\documentclass{article}

\usepackage{arxiv}

\usepackage[utf8]{inputenc} 
\usepackage[T1]{fontenc}    
\usepackage{hyperref}       
\usepackage{url}            
\usepackage{booktabs}       
\usepackage{amsfonts}       
\usepackage{nicefrac}       
\usepackage{microtype}      
\usepackage{lipsum}
\usepackage{algorithm}
\usepackage{algpseudocode}
\usepackage{amsmath}
\usepackage{graphicx}
\usepackage{multirow}
\usepackage{amsthm}
\usepackage[flushleft]{threeparttable}
\usepackage{adjustbox}
\usepackage{float}
\usepackage{verbatim} 
\usepackage{multirow}
\newtheorem{claim}{Claim}

\title{Ranking in Genealogy: Search Results Fusion at Ancestry}

\newcommand*\samethanks[1][\value{footnote}]{\footnotemark[#1]}
\author{
   Peng Jiang\thanks{Both authors contributed equally to this research.} \\
  Ancestry\\
  153 Townsend St., Ste. 800\\
  San Francisco\\
  California\\
  94107\\
  \texttt{pjiang@ancestry.com} \\
   \And
 Yingrui Yang\samethanks[1] \\
 Ancestry\\
  153 Townsend St., Ste. 800\\
  San Francisco\\
  California\\
  94107\\
  \texttt{yyang@ancestry.com} \\
 \And
  Gann Bierner \\
  Ancestry\\
  153 Townsend St., Ste. 800\\
  San Francisco\\
  California\\
  94107\\
  \texttt{gbierner@ancestry.com} \\
  \AND
  Fengjie Alex Li \\
  Ancestry\\
  153 Townsend St., Ste. 800\\
  San Francisco\\
  California\\
  94107\\
  \texttt{ali@ancestry.com} \\
  \And
  Ruhan Wang \\
  Ancestry\\
  153 Townsend St., Ste. 800\\
  San Francisco\\
  California\\
  94107\\
  \texttt{rwang@ancestry.com} \\
  \And
  Azadeh Moghtaderi \\
  Ancestry\\
  153 Townsend St., Ste. 800\\
  San Francisco\\
  California\\
  94107\\
  \texttt{amoghtaderi@ancestry.com} \\
}

\begin{document}
\maketitle

\begin{abstract}
Genealogy research is the study of family history using available resources such as historical records. Ancestry provides its customers with one of the world's largest online genealogical index with billions of records from a wide range of sources, including vital records such as birth and death certificates, census records, court and probate records among many others. Search at Ancestry aims to return relevant records from various record types, allowing our subscribers to build their family trees, research their family history, and make meaningful discoveries about their ancestors from diverse perspectives.

In a modern search engine designed for genealogical study, the appropriate ranking of search results to provide highly relevant information represents a daunting challenge. In particular, the disparity in historical records makes it inherently difficult to score records in an equitable fashion. Herein, we provide an overview of our solutions to overcome such record disparity problems in the Ancestry search engine. Specifically, we introduce customized coordinate ascent (customized CA) to speed up ranking within a specific record type. We then propose stochastic search (SS) that linearly combines ranked results federated across contents from various record types. Furthermore, we propose a novel information retrieval metric, normalized cumulative entropy (NCE), to measure the diversity of results. We demonstrate the effectiveness of these two algorithms in terms of relevance (by NDCG) and diversity (by NCE) if applicable in the offline experiments using real customer data at Ancestry.
\end{abstract}


\section{Introduction}
Ancestry provides its customers with a large-scale genealogical index with billions of records from a diverse range of sources, including vital records such as birth and death certificates, census records, court and probate records among many others. Within the database, users are able to search for historical records of their ancestors to construct their family trees.  A typical query consists of the name, gender, age and life events of the ancestor as well as his or her family members. 

Searching such a diverse set of content presents a number of challenges.  First, the types of data available from each source and their relevance to the query can vary greatly.  For example, while a married name is not present in birth records, it is valuable in census records and death records. Second, inexact matching of the query to documents is of central importance, as the names and places may have different synonyms over time. Additionally, inexact match should also be used for under-specified or slightly different date information, nearby or more generally specified places, as well as phonetically similar or misspelled names. Finally, some missing information needs to be inferred on the basis of the query. For example, missing data such as married names sometimes can be inferred.

Taken together, these issues make it particularly challenging for a search system to score records from such a diverse set of content. From the customer's perspective, the best possible results should be returned first with a satisfying diversity of result types. 

To address the widely found disparity issue in data sources, the Ancestry search system supports this ranking behavior by building a specialized query for each record type. For each record specific query, we apply a machine learning model 
to return a ranked list. This process is referred to as \textit{record specific search}. The results are then combined using a collator based on another machine learning model. This process is referred to as \textit{global search}.

For record specific search, we modify the weight initialization step in the coordinate ascent (CA) algorithm. This customized CA method is applicable to any application with binary features, achieving convergence 10 times faster relative to canonical CA in offline experiments while maintaining the ranking performance.

The global search is a special use case in federated search, where the databases have no overlaps and the relevance scores from each list are not necessarily cooperative. This is in contrast to classical federated search methods, where coexistence of documents and chorus effect are commonly exploited. On the other hand, linear combination models have proven effective in a number of studies to merge results from distributed information retrieval (IR) systems. Inspired by these models, we develop a new and practical heuristic stochastic algorithm (stochastic search, SS) that linearly combines the ranked lists. The number of ranked lists to be merged in our use case is around 8 to 20.  SS adopts Nelder$-$Mead algorithm with weights initialized by rankSVM, outperforming many state-of-the-art learning to rank models. Although heuristic algorithms optimizing for listwise loss are usually computationally expensive, their use in linear combination fusion problem with only 8 to 20 unknown parameters is feasible. Moreover, our approach directly optimizes for non-differentiable IR metrics such as normalized discounted cumulative gain (NDCG), thereby circumventing the problem of metric divergence. Our offline results show excellent computational efficiency as well as high relevance measure.

In addition to evaluating relevance in offline tests, it is important for us to return relevant records from diversified record types in order to assist our customers to better understand their family history. Ideally, the list of our search results should cover as many record types as possible at any position. A proper diversity metric can help evaluate the influence of record diversity on user engagements, thus allowing us to measure return investments when acquiring different types of record collections. Furthermore, we need the metric to measure not only global diversification, but also local diversification. Global diversification measures how many record types are present in the list, while local diversification measures whether the same or different record types are present between row to row records. For example, if different record types are represented by letters, such as A, B, etc, and $R_1^A$ represents that the first record is of type A, then the ranking list $L_1$ of [$R_1^A$, $R_2^A$, $R_3^B$, $R_4^B$] has the same global diversity as an another list $L_2$ of [$R_1^A$, $R_2^B$, $R_3^A$, $R_4^B$ ], but $L_2$ has better local diversification than $L_1$. 

To meet these requirements, inspired by how the ranking gain is accumulated and normalized in the definition of the ranking metric NDCG, we propose a new metric, which we term normalized cumulative entropy (NCE). NCE is based on Shannon entropy~\cite{Shannon:1949}. Krestel et al., 2012~\cite{Krestel2012} used entropy to measure diversity in their search results, demonstrating its utility in quantifying global diversity. To further measure the local diversity, we propose to sum up entropy value at each position. 
We divide the current value by a maximum cumulative entropy and generate the final metric score of NCE, which is in the range of 0 to 1. This allows us to compare the diversity of ranking lists with various lengths. The ideal cumulative entropy is defined as a special maximum entropy problem with an additional constraint that the probability of each type has to be a special value between 0 and 1 instead of any real value in that range. We formulate this problem as an integer programming maximization problem. To the best of our knowledge, this problem has never been studied before. We propose a way to find the optimal solution for our defined problem and use branch and bound algorithm to prove its correctness. As introduced in Section \ref{sec:diversity-review}, current existing diversity metrics are ineffective in measuring the diversity under our assumption. Our experiment on a toy data set showed that only NCE could successfully measure global and local diversity. Therefore, other than evaluating relevance by NDCG in our application, we also evaluate diversity by NCE. 

This paper is structured as follows. In Section \ref{background} we give a general overview of ranking problems in the context of genealogy. We then introduce the design of the Ancestry search engine architecture, followed by a description of queries and features used in the two models discussed in this paper. We also examine previous research work in federated search and diversity metrics. In Section \ref{model}, we describe the customized coordinate ascent model and its technical improvements, followed by a new heuristic algorithm that combines Nelder$-$Mead algorithm with rankSVM (i.e., stochastic search). In Section \ref{metric}, we propose a diversity metric for offline evaluation. In section \ref{experiments}, we show the relevance as well as the training time comparison on the real user search interaction data from our mobile app. 
\section{Background} \label{background}
\subsection{Ranking Problem in Genealogy}
At Ancestry, we apply learning to rank algorithms to a new area, genealogy, to assist our customers in better understanding their family history. The foundation of our service is an extensive and unique collection of billions of historical records that we have digitized, indexed and put online since 1996. We have the largest index of online records. Currently, our content collection includes 20 billion historical records. The record data consists of birth records, death records, marriage records, adoption records, census records, obituary records, among many others.  These records are digitized via optical character recognition (OCR) and indexed into the database over the past 20 years. These digital records and documents, combined with our proprietary online search technologies and tools, provide users with deeply meaningful insights about who they are and where they come from.

When users search within a large collection of records, they rely heavily on the search results to attach relevant records to ancestors in their family trees. This process allows them to further discover their family history by associating specific records with people in their family trees. To date, our customers have built 100 million family trees and 11 billion ancestor profiles on the Ancestry flagship site and its affiliated sites.

\subsection{Search Architecture at Ancestry}
In general, Ancestry search is similar to many other search systems.  A complicating factor lies in the fact that Ancestry provides genealogical content from a variety of sources, including vital records such as birth and death certificates, census records, court and probate records.  When searching for a person within these records, the search system must take into account the fact that search terms may be more valuable within some types of records than in others.  

\begin{figure}[h]
  \centering
  \includegraphics[width=0.6\linewidth]{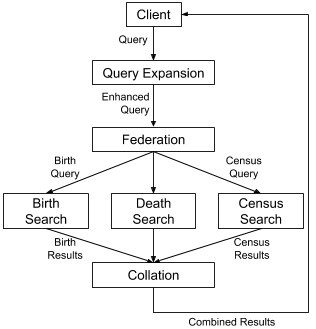}
  \caption{Example Search Process}
  \label{fig:architecture}
\end{figure}

To tackle this problem, Ancestry processes queries in a few basic stages.
\begin{enumerate}
    \item The query is received from the client. It may contain a variety of information to search for, including, for example, given and surnames, life events (dates and places), and names of family members.
    \item The query is expanded and enhanced based on the specific characters of record types. For example, similar names may be included, places and dates may be normalized.
    \item The expanded query is distributed to search servers, which process the query and return a list of matching records from each record type. This stage is referred to as the record specific search.
    \item Finally, in global search, a collator combines the records into a single ranked list, which are then returned to the client.
\end{enumerate}
We illustrate the procedure by an example in Figure~\ref{fig:architecture}.

\subsection{Queries and Features}
Queries are typically a partial description of a person and may include information such as name, gender, relatives, and dates and places for life events (e.g., birth and death). The desired search results are records in which this person is present.

Queries may be noisy due to typos or misinformation. It is common, for example, for a query to be off by a year or two in a birth date, or to misuse a middle name for a first name. Records can also be noisy due to the OCR and keying errors. Because of the noise in both the queries and records, a certain amount of fuzziness must be allowed to achieve acceptable recall. 

In addition, we can infer useful information from a given query. For example, if the birth date is absent from the query, it can be inferred from death date and age at death.

Taken these factors into consideration, a query expansion module translates the facts expressed in the query into a more extensive set of search clauses. These clauses may apply to a variety of life events (birth, death, marriage, residence, etc) and people (self, spouse, parents, children, etc), yielding hundreds of potential clause types with different information. In record specific search, we define binary features to indicate whether a particular clause in the query matches the clause in the record. We employ machine learning models based on these features for record specific search and predict a score for each record in the ranked list.

In global search, features are the predicted score from record specific search and the machine learning model will learn a weight for each ranked list of a given record type. The final ranking score of each record is then re-scaled by its corresponding record type weight. The ranking of the global search results is determined by the descending order of the final ranking score.

\subsection{Related Work}
\subsubsection{Federated Search} \label{fusionreview}

In realistic scenarios distributed information retrieval (IR) has gained increasing attention. It is common practice to merge the results of multiple queries to an optimal rank list. Previous work can be classified into two major use cases. In collection fusion, document scores returned by different databases are uncooperative. Most of the work focused on the use case that the same documents can appear in different databases / result lists. In \cite{shokouhi_robust_2009} a global score based on co-existence of a document in different queries in proposed. RRF \cite{palakodety_query_nodate} weighted the importance of a document based on its rank in the original query and number of occurrence of the document (inverse squared rank). ICTNET \cite{guan_ictnet_nodate} used page rank to evaluate importance of a document, and calculated similarity between query representations and the document, using a weight similar to RRF to weight the similarity.

The other line of research work focuses on data fusion, in which a query is issued to multiple retrieval models that have access to the same document collection. Chorus effect indicates that usually the merged list will have greater relevance than the output of any single model. ProbFuse \cite{lillis_probfuse:_2006} merged the results from multiple queries on the same data collection by analyzing the performance of each individual model and the probability of the relevance of documents returned. LambdaMerge \cite{sheldon_lambdamerge:_2011} merged results from reformations of a query with a gated neural network based score combination function that utilized multiple query-document features. 

Based on the nature of the design of search architecture at Ancestry, our use case is a problem combining the previous use cases as the databases don't share common documents and different ranked list can be trained by different machine learning models. 

Linear combination models have been proposed in a number of studies. Some of the approaches require additional information about the documents and the system while some can only apply to certain types of fusion problems. Some supervised ensemble learning approaches can combine ranked lists predicted by multiple models into a single ranked list, among which, \cite{wu_adapting_2010,burges_ranknet_nodate} combined the tree based learning with rank (LtR) methods (LambdaMART, lambdaRank, MART) trained on variations of the training set and achieved performance better than using a single model. They found that combining normalized score by weight 1 provided similar results as those using optimal combiner, which combined pairs of ranks one by one as used in combining weak learners in LambdaMART. It is also similar to those combiners using LtR models (lambdaRank etc.).  Vogt \cite{vogt_fusion_1999} derived a formula for a theoretically optimal weighting scheme for combining two systems. Si and Callan \cite{si_using_2002} used three different types of information to normalize the document scores: resource description, a database score, and the document score returned by the database. Powell et al. \cite{powell_impact_2000} compared raw-score merging and normalized score merging. Cossock and Zhang \cite{cossock_subset_2006} focused on top results in ranked lists and proposed an approximate minimization of certain regression errors showing it approximately optimized for NDCG at top positions. Larkey et al. \cite{larkey_collection_2000} found that for rank merging, normalized scores are not as good as global IDFs for merging when the collections are topically organized. None of these methods focus on combining multiple ranked lists using only predicted scores from each list. Additionally, they don't directly optimize for IR metrics such as NDCG. 

\subsubsection{Existing Diversity Metrics} \label{sec:diversity-review}

Diversity has attracted great interests from the machine learning community. Researchers have proposed several popular metrics including $\alpha$-NDCG \cite{Clarke08}, and a family of Intent-Aware (IA) metrics \cite{Agrawal09}, including NDCG$-$IA, MRR$-$IA, and MAP$-$IA.

However, several limitations hamper the widespread application of these previously developed metrics. First, it is hard to evaluate different results based solely on diversity. Many diversity metrics measure relevance and diversity at the same time, or account for both novelty and diversity in search ranking results. Thus, the difference of two results might be dictated by either relevance or novelty. Second, many intent-aware metrics require information about user intents. Unfortunately, it is often nontrivial to obtain these information. Last, for many short or ambiguous queries, user's interest is unclear or even unspecified~\cite{Radlinski06}.

We aim to develop a metric to evaluate search diversity independently. The metric should not be mixed with other metrics, such as relevance or novelty, nor should it depend on any extra information such as user intents. S-recall \cite{zhai03} meets these requirements. It is used to measure the diversity of subtopic retrieval problem aiming to find documents that cover many subtopics. However, this metric could only evaluate global diversity but not local diversity.

\section{Models} \label{model}
\subsection{Customized CA for Record Specific Search}
CA is an optimization algorithm similar to gradient descent. It first initializes all variables randomly. Then during each iteration it alternatively updates a variable by a step size along a certain direction that leads to the local minimum or maximum of the objective function. CA has a similar convergence property as gradient descent. However, a big difference between them is that CA is a derivative-free optimization algorithm. That is to say, CA could optimize an objective function for which the derivative information is not available. This works well for ranking problem, as the objective function is a evaluation metric such as NDCG, which could be represented as a non-differentiable function in terms of feature weights.            

However, CA is known for its slow convergence. We propose an weight initialization schema that makes use of the labels in the training data for applications with binary feature values. 

In the CA implementation provided by the library RankLib~\cite{dang_lemur_nodate}, it initializes the same weight for every feature as 1 divided by the number of features. 
In record specific search, the value of a feature is either 1 or 0, and the label is either relevant or irrelevant. Assuming a feature's value is 1 in only relevant records, it makes sense to initialize a higher weight to this feature. Then when calculating the weighted feature sum as the predictive score for each record, the relevant records will be predicted a higher score than irrelevant ones, therefore be ranked on top of the list. This help increase the objective function such as NDCG, thus speed up convergence time to improve NDCG. 

Specifically, we propose to initialize a weight of each feature as follows:

\begin{equation}
w = 
\begin{cases}
    0.5,& \text{if $fre\_rel = 0$, $fre\_irrel = 0$ }\\
    \frac{fre\_rel}{fre\_rel + fre\_irrel} & \text{otherwise }
  \end{cases}
\end{equation}

where $fre\_rel$ is the number of times for the feature being 1 in relevant records, and $fre\_irrel$ is the number of times for the feature being 1 in irrelevant records. 
 
Next we use a toy data to show how to initialize the weights. Table~\ref{toy_data} gives the label and feature values for 4 record. A label with value 1 means that the record is relevant, and a label with value 0 means that the record is irrelevant. 
\begin{table}{}
\centering
\caption{Toy Training Data}
\begin{tabular}{| c|c ||c c c c|} \hline
record & label & $f_1$ & $f_2$  & $f_3$ & $f_4$\\ \hline
$r_1$ & 1  & 1   & 0  & 1 & 0 \\ \hline
$r_2$ &1  &  1 &  0 & 1 & 1\\ \hline
$r_3$ &0  &   0 &  1 & 1 &  1 \\ \hline
$r_4$ &0  &    0&   0&  0 & 0 \\ 
\hline\end{tabular}
\label{toy_data}
\end{table}
We could then calculate the initialized weight of each feature, indicated by $w_{new}$ in Table~\ref{feature_data}. For example, $f_3$ is 1 in relevant records for 2 times, and is 1 in irrelevant records for 1 time, thus the initial weight is 2/(2+1), which is 2/3. For comparison, the weights obtained by default CA would be 1 divided by the number of features, i.e., 1/4 for each feature, indicated by $w_{old}$.
\begin{table}{}
\centering
\caption{Feature Weights}
\begin{tabular}{|c ||c c c c|} \hline
 & $f_1$ & $f_2$  & $f_3$ & $f_4$\\ \hline
$w_{new}$  &   1 &  0 & 2/3 &  0.5 \\ \hline
$w_{old}$  &   1/4 &  1/4 & 1/4 &  1/4 \\
\hline\end{tabular}
\label{feature_data}
\end{table}

Notice in this way the weight is in the range of 0 and 1. It is trivial to prove that if a set of weights is not in this range, then by scaling them  will change only the absolute predicted scores of records, but the relative order will remain the same, therefore leading to the same ranking and NDCG.

Now the question is whether $w_{new}$ could improve NDCG when compared with $w_{old}$. To answer that question, we  calculate the score for each record based on two sets of feature weights. Specifically, a score is predicted for a record as the linear combination of its features and weights. For example, from Table~\ref{toy_data} we know that the $r_1$ has features $[1, 0, 1, 0]$, using $w_{new}$ of $[1, 0, 2/3, 0.5]$, we predict a score of 1.67 for $r_1$, as $1*1+0*8+1*2/3+0*0.5 = 1.67$. 

Table~\ref{rank} shows the score for each record. $s_{new}$ and $s_{old}$ stands for the score predicted by customized CA and default CA respectively.
 $rank_{new}$ is the rank obtained by customized CA. And the label of each record is copied from Table~\ref{toy_data}. Based on the descending order of the scores in $s_{new}$, customized CA puts both relevant records on top of the list, therefore leading to a optimal NDCG score of 1. On contrast,  the default CA would predict a lower score to $r_1$ than that of $r_3$, therefore ranking a irrelevant record on top of a relevant one. 

\begin{table}{}
\centering
\caption{Score and Rank}
\begin{tabular}{| c|c ||c || c|} \hline
record & label & $s_{new}$  &  $s_{old}$  \\ \hline
$r_1$ & 1  & 1.67 &  0.5  \\ \hline
$r_2$ &1  &  2.17  &  0.75 \\ \hline
$r_3$ &0  &   1.17 & 0.75   \\ \hline
$r_4$ &0  &    0 &   0  \\ 
\hline\end{tabular}
\label{rank}
\end{table}

We experimentally show in Section~\ref{sec:ca_expe} that customized CA help speed up the procedure 10 times faster in record specific search. 

\subsection{Stochastic Search (SS) for Global Search}
 As mentioned in the previous section, the problem is in essence a numerical problem with $N$ unknown weights to linearly combine multiple ranked lists into one, targeting a non-differentiable objective function. Usually the dimension of the parameter space $N$ is moderate. We adopt a simplified approach using Nelder$-$Mead (Downhill Simplex) method \cite{mckinnon_convergence_1998}. It achieves fast convergence and enjoys an improvement in NDCG compared to other state-of-the-art models. 

The Nelder$-$Mead method is a classical numerical method to find the optimum of a loss function in a multidimensional space. It is a heuristic search method that updates downhill through a multidimensional objective that can converge to non-stationary points \cite{powell_search_1973}.  

The initial weights play a crucial role in the speed of convergence as well as the performance of the algorithm if there are multiple local optimums. 
Instead of random initialization, some prior knowledge about the possible range of the weights could greatly speed up the process and ensure a good performance. Along these lines, we take a small sample from the training data to learn the initial weights using a simple linear model such as rankSVM, which are then used as the starting values for the Nelder$-$Mead algorithm. We refer to the this two-step approach as stochastic search (SS). 

In this section, the rank fusion problem is an optimization problem with a loss function $L(y, \boldsymbol{w})$, where $\boldsymbol{w} = [w_1, w_2, ..., w_N]$ is an $N$ dimensional vector representing unknown weights of $N$ ranked lists; $y$ is the target. Each ranked list $R_n$ consists of $d_n$ documents with a ranking score $s_{ni}$. The list of records returned to users ($\bar{R}$) is composed of a linear combination of the ranked list $R_1,...,R_N$. the predicted ranked score $\bar{s}_i = \sum_{j=1...N} w_j s_{ji}$ is used to rank the combined ranked results $\bar{R}$. Note that the loss function is defined as an IR metric calculated based on $\bar{R}$, such as NDCG@k. 

\begin{figure}[h]
  \centering
  \includegraphics[width=0.6\linewidth]{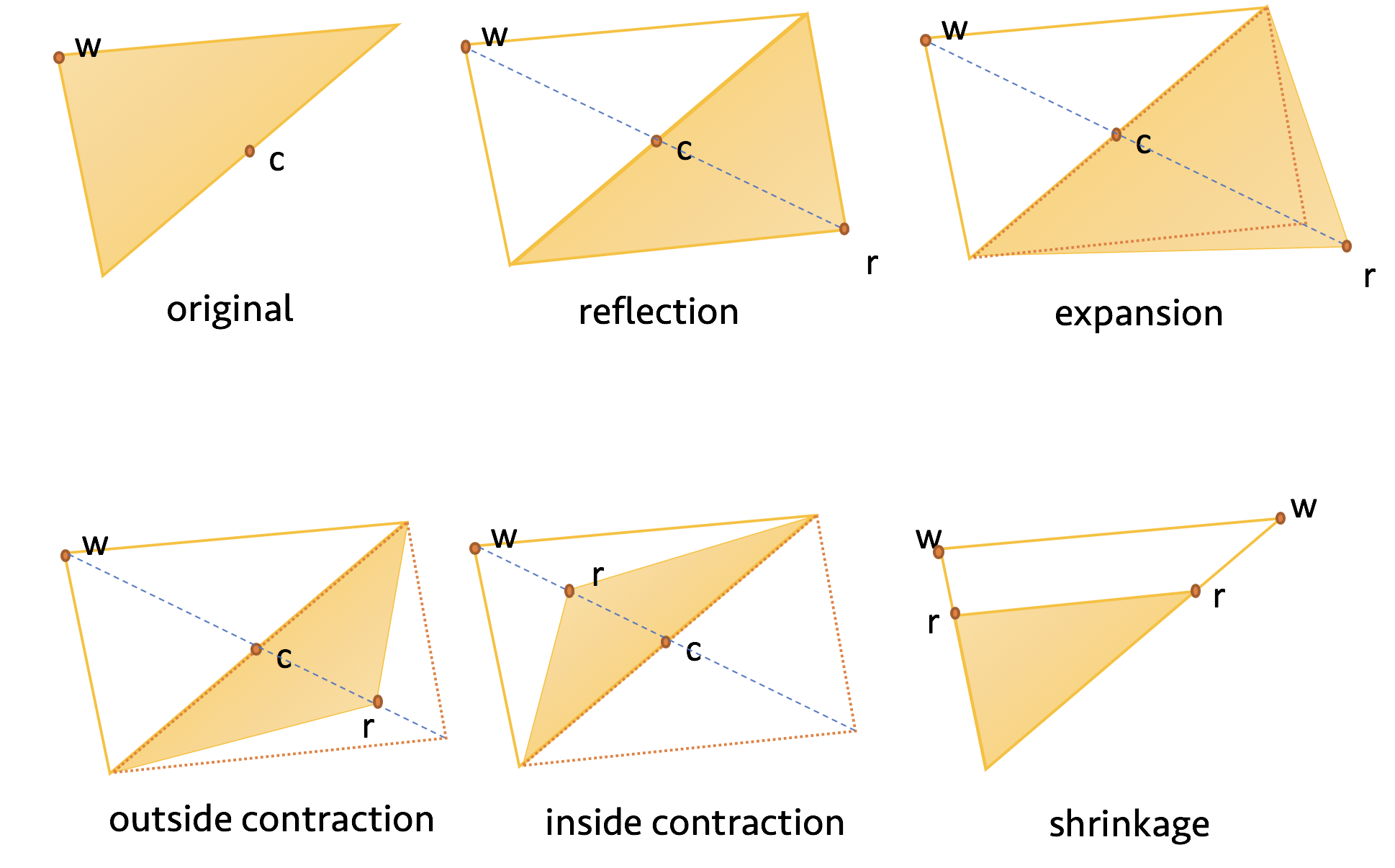}
  \caption{Operations in the Nelder$-$Mead simplex algorithm}
  \label{fig:simplex}
\end{figure}

\begin{figure}[h]
  \centering
  \includegraphics[width=0.6\linewidth]{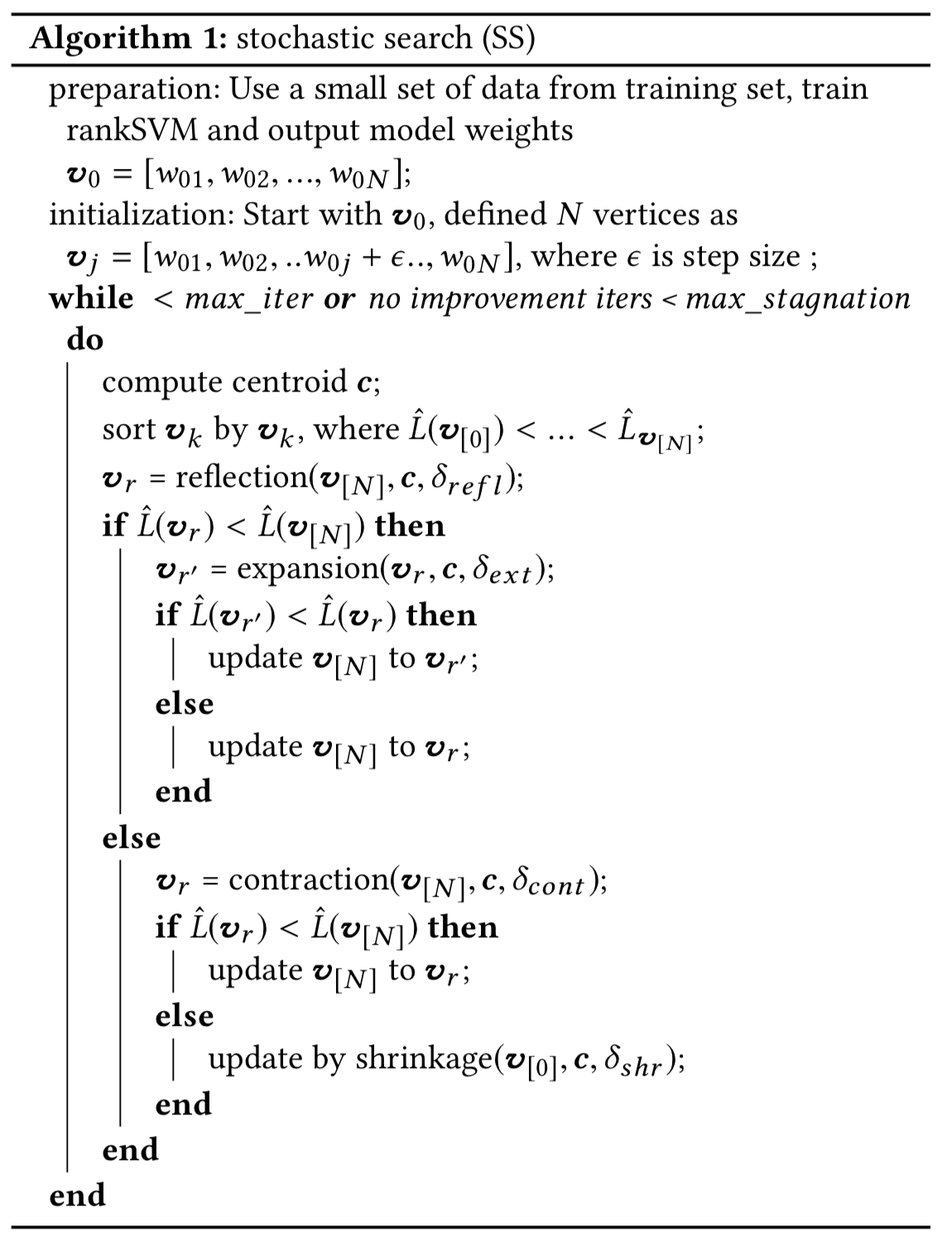}
  \caption{Stochastic Search Algorithm}
  \label{fig:pa}
\end{figure}

To initialize the algorithm, a point 
$\boldsymbol{v}_0 = [w_{01}, w_{02}, ..., w_{0N}]$ with a step size $\epsilon$ is chosen. The initial simplex of vertices consists of the initial point $\boldsymbol{v}_0$ and $\boldsymbol{v}_j$ where $\boldsymbol{v}_j = [w_{01}, w_{02}, ..w_{0j}+\epsilon.., w_{0N}]$, $j=1,...N$. Initial value of loss function $\hat{L}_0$ is evaluated at all $N+1$ vertices. At each iteration, the vertices ($\boldsymbol{v}_k$) are ordered by $\hat{L}(\boldsymbol{v}_k)$, then the parameter space is explored by operations including reflection, expansion, contraction and shrinkage. Reflection moves the vertex with highest $\hat{L}(\boldsymbol{v}_k)$ to another point to reduce the loss function, while preserving the volume of the simplex. If the new vertex produces smaller loss function, the algorithm expands the volume by extending the vertex, otherwise it contracts itself in the transverse direction. If contraction doesn't work, it will shrink itself in all directions around the best vertex (with lowest objective function). Figure~\ref{fig:simplex} shows the definition of each of the operations and Algorithm~\ref{fig:pa} shows the steps of SS.

Nelder$-$Mead method explores the local area of a parameter combination within each step and updates multiple parameters accordingly in each iteration. It prevents the algorithm from going too greedily into a local optimal.  The step size $\epsilon$ of the change in weights must be big enough to allow some exploration of the space. $max\_iter$ determines the maximum number of iterations of the algorithm while $max\_stagnation$ determine the maximum number to try when there's no improvement in the loss function.  $\delta_{refl}, \delta_{ext}, \delta_{cont}$ and $\delta_{shr}$ determines the ratio of changes in the operations.  $\delta_{refl}=1$ is a standard reflection, $\delta_{ext}$ represents the amount of the expansion; 0 means no expansion. $\delta_{cont}$ is the contraction parameter and should be between 0 and 1. $\delta_{shr}$ is the ratio of shrinkage, should be 0 to 1. 

If the rate of convergence of the algorithm is too low or the algorithm often stuck into local optimum, we could combine SS with other approaches to increase stochasticity across iterations. For example, NDCG-Annealing method \cite{karimzadehgan_stochastic_2011-1} combines downhill simplex algorithm with Simulated Annealing algorithm that can help with this issue. Similarly, we could apply SS at each iteration and set a temperature for the probability of updating the cost function based on simulated annealing method. 

\section{Evaluation Metrics} \label{metric}

 We use normalized discounted cumulative gain (NDCG) as the primary metric, which has been widely used to assess relevance in ranking problems. Let $x_1,...x_n$  be the $n$ documents in the ranked list. Let $Y$ be the set of degrees of relevance. In our use case, $Y = \{0, 1\}$ where $0$ corresponds to irrelevance and 1 to relevance. Let $y_1$,..., $y_n$ ($y_i \in Y$) be the degree of relevancy associated with $x_1$,...,$x_n$.  Let $f$ be a ranking function that maps each document to a ranking score $f(x_i)$. The resulting ranked list, denoted by $x^f_{(1)},...,x^f_{(n)}$ satisfies $f(x^f_{(1)}) > ... > f(x^f_{(n)})$, where $y^f_{(1)},...,y^f_{(n)}$ are the corresponding relevance. DCG@k is defined as $$DCG@k=\sum_{i=1}^k \frac{y^f_{(i)}}{log_2(1+i)}$$,
 and IDCG@k defined as the DCG value of the best ranking function on the ranked list $$IDCG@k = max_{f'}\sum_{i=1}^k \frac{y^{f'}_{(i)}}{log_2(1+i)}$$. NDCG@k is defined as $$NDCG@k=\frac{DCG@k}{IDCG@k}$$.

Other than using NDCG to measure ranking relevance, our global search application desires a proper diversity metric to measure influence of diversity of record collections on user engagements. To measure the diversity under our assumption that an optimal diversified ranking list should present as many record types as possible at any position, we propose a diversity metric normalized cumulative entropy (NCE) to measure both global and local diversity.

Inspired by how the ranking gain is accumulated and normalized in the definition of the ranking metric NDCG, NCE measures the diversity of a given ranking list in three steps. First, we use the popular Shannon entropy formula to calculate the entropy. This could successfully evaluate the global diversity. Second, we propose to sum up entropy value at each position. That is where the cumulative entropy comes from. Cumulative entropy could successfully evaluate the local diversity. Third, we divide the current value by an ideal cumulative entropy and generate the final metric score of NCE, which is in range between 0 and 1. This makes it possible to compare diversity for ranking lists with different lengths. The ideal cumulative entropy is defined as a special maximum entropy problem with an additional constraint that the probability of each type has to be a special value between 0 and 1 instead of any real value in that range. We formulate this problem as an integer programming maximization problem. To the best of our knowledge, this problem has never been studied before. We propose a way to find the optimal solution for our defined problem and use branch and bound algorithm to prove its correctness.

We first introduce how entropy is applied to measure global diversity.
Given an ordered list $Q$ of $n$ documents $\{d_1, d_2, ..., d_n\}$. Each document belongs to one record type and there are $K$ record types in total. 
The Shannon entropy for this list $Q$ is defined as follows:
\begin{equation}
E(Q) = -\sum_{i=1}^{K}p_i\log{p_i}, 
\label{entropy}
\end{equation}
where $p_i$ stands for the probability of record type i. It is calculated as follows:
\begin{equation}
p_i = \frac{n_i}{n},
\label{prob}
\end{equation}
where $n_i$ stands for the number of documents belonging to record type $i$, and $n$ is the total number of documents.

For example, consider an ordered list \textit{AAAB} which has four records belonging two categories A and B, where A and B stand for any record type in our application. For convenience here we ignore the record number and show only the record types. Then the entropy for this list is:
\begin{equation*}
E(AAAB) =-\frac{3}{4}\log\frac{3}{4}-\frac{1}{4}\log\frac{1}{4} = 0.811
\end{equation*}

Using entropy will tell an ordered list \textit{AABB} has better global diversity than \textit{AAAB}. However, it could not differentiate the local diversity between \textit{AABB} and \textit{ABAB}. 

To address this issue, we propose to sum up entropy value at each position and name it cumulative entropy. 
\begin{equation}
CE(Q) = \sum_{p=1}^{n}E(Q_p)
\label{ce}
\end{equation}
where $E(Q_p)$ is the entropy defined in~(\ref{entropy}) for list $\{d_1, d_2, ..., d_p\}$.

For example, the cumulative entropy for list \textit{AABB} is:
\begin{equation*}
CE(AABB) = E(A)+E(AA)+E(AAB)+E(AABB) = 0 + 0 + 0.918 + 1 = 1.918
\end{equation*}
Similarly, we could calculate the cumulative entropy for list \textit{ABAB}:
CE(ABAB) = E(A)+E(AB)+E(ABA)+E(ABAB) = 0 + 1+ 0.918 + 1 = 2.918
Thus cumulative entropy could evaluate local diversity successfully.

However, in our global search where the result list varies in length for each query, the comparison of cumulative entropy would not make sense across queries. Thus we propose to divide the current value by an ideal cumulative entropy and name the metric normalized cumulative entropy (NCE), which is in range between 0 to 1. NCE makes it possible to compare diversity for ranking lists with different length.

Now the key issue is how to define the ideal or maximum entropy at each position. It is well known that entropy is maximized when the probability distribution is uniform, but this conclusion does not hold in our application. For example, assume there are 3 record types and we are interested in maximum entropy of top 5 documents, then what should be the probability of each record type to achieve maximum entropy? If the probability distribution is uniform, i.e., 
each record type has probability 1/3. Then this indicates that in top 5 documents, there are $\frac{5}{3}$ documents belonging to each type. This does not make sense in our application. 

We formulate this problem as an integer programming maximization problem. Specifically, the maximum entropy of K record types at top n documents could be formulated as follows:
\begin{gather*}
	\max \quad  - \sum_{i=1}^{K}\frac{n_i}{n}\log{\frac{n_i}{n}}  \\
	\begin{aligned}
	\textup{s.t.}\quad &\sum_{i=1}^{K}n_i =  n\\
	&n_i \in \{0,1,2,...,n\}\\
	\end{aligned}
	\label{ie_integer}
	\end{gather*}
where $n_i$ stands for the number of documents belonging to record type $i$. Recall that $p_i$ stands for the probability of each record type. By substituting $n_i$ with $p_i$ according to Equation~\ref{prob}, the above formulation could be represented in terms of $p_i$ as follows:
\begin{gather}
	\max \quad  - \sum_{i=1}^{K}p_i\log{p_i}  \\
	\begin{aligned}
	\textup{s.t.}\quad &\sum_{i=1}^{K}p_i =  1 \nonumber \\
	&p_i \in \{0,\frac{1}{n},\frac{2}{n},...,\frac{n-1}{n}, 1\}\\
	\end{aligned}
\label{ie}
\end{gather}
The problem is NP-hard. There is a strong constraint on $p_i$ requiring it to be a special value between 0 and 1 instead of any real values in that range. If we relax this constraint to allow $p_i$ take any real value between 0 and 1, then the relaxed problem is the popular well-studied maximum entropy problem. 

To the best of our knowledge, the problem formulated in~(\ref{ie}) has never been studied before. Next we propose a way to find its optimal solution and use branch and bound algorithm to evaluate its correctness.

The optimal solution of ~(\ref{ie}) is stated in CLAIM~\ref{ieproof2}.
\begin{claim}
	For K record types, the maximum entropy value of top n documents happens when there are (n/K) documents belonging to each of (K -$n\mod K$)  record types. If ($n\mod K $) is 0, we are done. Otherwise for the ($n\mod K$) record types, there are (n/K)+1 documents belonging to each of these record types. 
	\label{ieproof2}
\end{claim}	
Go back to the previous example where we assume 3 record types and are interested in maximum entropy of top 5 documents, we first calculate $5/3 = 1$, and $5 /mod 3 = 2$. Then according to Claim~\ref{ieproof2}, we know the maximum entropy happens where there are 2 documents belonging to each of the two record types, and there are 1 document belonging to the remaining one record type. Thus the maximum entropy happens when the probability of each record types will be $(2/5,2/5,1/5)$ instead of $(1/3,1/3,1/3)$. 

To evaluate CLAIM~\ref{ieproof2} leads to the maximum entropy for problem~(\ref{ie}), we could always use brute force method to compare  limited number of feasible solutions, and verify whether the optimal solution satisfies the conclusion in CLAIM~\ref{ieproof2}. However, there are many feasible solutions which we know will not improve entropy value without evaluating them. Thus we 
we apply a branch and bound algorithm to cut off branches where solutions  are unnecessary to be evaluated. For details, please refer to Appendix. 

\section{Experimental Results} \label{experiments} 
\subsection{Customized CA for Record Specific Search}
\label{sec:ca_expe}
In this section we compared the performance and running time of default CA and customized CA using real customer data from record specific search. The purpose was to check whether customized CA could reduce running time while maintaining performance. We randomly chose marriage record and residence record in our example for illustration. Table~\ref{stats_eval} lists the number of samples and features in each record type.  
\begin{table}
\centering
\caption{Statistics of  Records}
\begin{tabular}{| c|c|c |} \cline{2-3}
 \multicolumn{1} {c|}{} & \# of samples & \# of features   \\ \hline
 marriage & 60232  &   121 \\  \hline 
 residence & 69053   & 87  \\   
\hline\end{tabular}
\label{stats_eval}
\end{table}

The relevance performance and running time is shown in Table~\ref{table:ca-perf}. The results on either marriage record or residence record supported the conclusion that customized CA had similar ranking performance as default CA but run approximately 10 times faster. The speed up in training procedure would then allow us to have a faster iteration with new training data, new feature set and new partitioning scheme.

\begin{table}
\centering
\caption{Comparison of NDCG and time between CA and Customized CA}
\begin{tabular}{|c|c|c|c|c|}
\hline
\multirow{2}{*} {Record Type} & \multicolumn{2}{c|}{NDCG }& \multicolumn{2}{c|} {Time (s)} \\ \cline{2-5} 
                          & {CA}       & {Customized CA}        & {CA}         & {Customized CA}         \\ \hline
{marriage}   & 0.5567       & 0.5534        &    715     & 68       \\ \hline
{residence} &  0.5966 & 0.5918        & 767        & 72       \\ \hline
\end{tabular}
\label{table:ca-perf}
\end{table}

\subsection{Diversity Metric Comparison on Toy Data}
\label{sec:nce_expe}
In this section we compared NCE with several existing diversity metrics on a toy data set with three ranking lists. Table~\ref{table:comp} shows the data information and the evaluations by different metrics. Each letter such as A and B, stands for a unique record type. 

Under our assumption that an optimal diversity list should cover as many record types as possible at any position, we observed that the second ranking list had the optimal diversity. 
Table~\ref{table:comp} shows that only S-recall and NCE correctly showed that the second ranking list had a better diversity than the first one. It is within expectation that NDCG-IA, MRR-IA, and MAP-IA were not able to correctly measure diversity, as they were designed to measure diversity and relevance simultaneously.

Furthermore, S-recall could not differentiate local diversity between the second and third ranking list, as it counted the total number of different record types yet ignored how they appeared at different positions. We concluded that only NCE successfully evaluated the global and local diversity. Therefore, we adopted NCE to evaluate diversity in our global search application.

\begin{table}
\centering
	\caption{Comparison of popular diversity metrics}
\resizebox{.46\textwidth}{!} {%
	\begin{tabular}{ccccc}

		\toprule
		Position & Ranking list 1& Ranking list 2 &Ranking list 3 & Relevant Label\\
		\midrule

		1 & A & A & A&1\\
		2 & A & B & A&0\\
		3 & B & C & B&0 \\
		4 & B & D & B&1\\
		5 & B & A & C&1\\
		6 & C & B & C&0 \\
		7 & C & C & D&1\\
		8 & C & D & D&0\\

		\midrule
		NDCG-IA@8 & 0.775 & 0.658  & 0.908 \\
		\midrule
		MRR-IA@8 & 0.667 & 0.625 & 0.875\\
		\midrule
		MAP-IA@8 &  0.694 & 0.625&  0.875\\
		\midrule
		S-recall@8 & 0.750 & 1.000 & 1.000\\
		\midrule
		NCE@8 & 0.025 & 0.041& 0.030\\
		\bottomrule
	\end{tabular}%
}
\label{table:comp}
\end{table}

\subsection{Performance Evaluation of Stochastic Search for Global Search}
We compared the performance of SS with popular learning to rank algorithms on a sample of global search data from the Ancestry mobile app. We note that in this use case, customized CA could not be applied as the feature values are not binary. We had in total around 30,000 queries and randomly partitioned them into three folds of training/testing data. The following experiments are reported using the averaged results across the three folds. We used NDCG as the primary metric to measure relevance, and NCE as the secondary metric to measure diversity. We also compared the running time of each algorithm.

We divided user's interactions into two classes of labels: negative (no interaction) and positive (if the user attaches the result to the family tree). Queries without any positive interactions were removed from the training and testing set. 
A query was expanded and customized based on specific requirements for each record type. A record specific machine learning model is then applied to retrieve top 100 (if applicable) records from each record type. The length of returned ranked list from different record types were not always 100, because some record types don't have enough matched results to return. Each record then had a predicted score by the record specific model. These predicted scores were features we fed into global search machine learning models discussed in this section, which predicted a weight for each record type. The final ranking score of each record was then rescaled by its corresponding record type weight. A collator combined the results from these lists into a single ranked list based on the descending order of the final ranking score.

In Table~\ref{tab:ndcg}, we compared SS with classical models implemented in RankLib (LambdaMART, RankNet, RankBoost, AdaRank, ListNet, CA) \cite{dang_lemur_nodate} and rankSVM model \cite{chang_libsvm:_2011} for their ranking relevance. The baseline was the raw-score combination in which we combined results across record types based on predicted scores from each record specific model. We defined the objective function of SS to be NDCG@100, since NDCG@100 was the primary metric of our offline experiments. The size of the subset used to learn the initial weights of SS was around 1$,$000 queries. The step size $\epsilon$ was 0.1; $max\_stagnation$ was set to be 10; $\delta_{refl}$ was 1, $\delta_{ext}$ was 2; $\delta_{cont}$ was 0.5 and $\delta_{shr}$ was 0.5. We observed that complex models such as RankNet and AdaRank suffered from over-fitting. Simple models such as RankSVM were more effective in optimizing NDCG in this case. 
Among all the models, SS provided the best performance. It outperformed rankSVM significantly, indicating that Nelder-Mead steps played a critical role in finding an optimal solution for NDCG@100. The performance of SS was expected as it optimized NDCG@100 directly. 

\begin{table}
\centering
  \caption{NDCG@100 on test datasets for different models.}
  \label{tab:ndcg}
  \begin{tabular}{ccc}
    \toprule
    methods & NDCG@100 & +/- baseline (\%)\\
    \midrule
    baseline & 0.6072 & -- \\
    LambdaMART & 0.7106 & 10.34\\
    RankNet & 0.2918 & -31.55\\
    AdaRank & 0.5865 & -2.07\\
    ListNet & 0.6619 & 5.47\\
    RankSVM & 0.7050 & 9.78\\
    RandomForest & 0.7029 & 9.57\\
    MART & 0.7001 & 9.29\\
    CA & 0.7107 & 10.35\\
    SS & 0.7131 & 10.58\\
  \bottomrule
\end{tabular}
\end{table}

We also applied NCE as the secondary metric to measure diversity. The NCE@100 of the models was shown in Table~\ref{tab:nce}, indicating that LambdaMART, RankSVM, CA and the SS model improved NCE@100 compared to baseline. Among them, LambdaMART, CA and SS all had more than 1\% improvement in NCE compared to baseline,  while RankSVM have around 0.38\% improvement. As shown in the NCE evaluation on ranking list 2 and 3 in Table~\ref{table:comp}, it was found that 1\% improvement of NCE value indicated a substantial improvement over the diversity in the results.

\begin{table}
\centering
  \caption{NCE@100 on test datasets for different models.}
  \label{tab:nce}
  \begin{tabular}{ccc}
    \toprule
    methods & NCE@100 & +/- baseline (\%)\\
    \midrule
    baseline & 0.1859 & -- \\
    LambdaMART & 0.2000 & 1.41\\
    RankNet & 0.0111 & -17.48\\
    AdaRank &0.0809 & -10.5\\
    ListNet & 0.1437 & -4.22\\
    RankSVM & 0.1897 & 0.38\\
    RandomForest & 0.1711 & -1.48 \\
    MART & 0.1774 & -0.85\\
    CA & 0.2004 & 1.45 \\
    SS &  0.1995 & 1.36\\
  \bottomrule
\end{tabular}
\end{table}

Based on the result from NCCG@100 and NCE@100, we concluded that CA, SS, LambdaMART were the most effective models regarding ranking relevance and diversity. To further differentiate the performance among CA, SS, LambdaMART, we compared their running time in Table~\ref{tab:time}. Each model was trained using an AWS EC2 instance with 8 vCPU and 61GB memory. LambdaMART took a long time to finish the training process, while SS converged very rapidly. Furthermore, the running time of SS was more than 2 times faster than that of CA. 

In summary, SS represents a highly effecitive method in terms of both ranking relevance and diversity. Moreover, it requires a minimal amount of training time, allowing faster iteration with new training data, new feature set, and new partitioning scheme.


\begin{table}
\centering
  \caption{Training time for different models.}
  \label{tab:time}
  \begin{tabular}{cc}
    \toprule
    methods & Training Time (s)\\
    \midrule
    LambdaMART & 14,021\\
    CA & 2604\\
    SS & 1172\\
  \bottomrule
\end{tabular}
\end{table}

\section{Conclusion}
In this paper, we described two heuristic stochastic models, including the customized CA model and the stochastic search model. Customized CA can be used in applications with binary features. It is shown to accelerate convergence with a 10 fold improvement compared to traditional coordinate ascent in record type specific search data. SS enables rapid convergence in federated search, where 8 to 20 ranked lists need to be merged.

Furthermore, our global search application demands a new diversity metric to evaluate the impact of record collection diversity on user engagements, thus allowing our company to measure return on investment (ROI) when acquiring different types of record collections. We proposed a new metric (normalized cumulative entropy, NCE) to determine the diversity of returned results from different record types, assuming there is no preference for a particular result type at any position. Offline experiments on toy data show that only NCE could successfully measure both global and local diversity among several popular existing diversity metrics. 

Based on these findings, in our offline experiments with real customer data, we use NDCG as our primary metric and NCE as the secondary metric to evaluate the performance of both relevance and diversity among popular learning to rank algorithms. Offline experiments show that stochastic search provides the best result in terms of NDCG. Additionally, NCE and training time-based analysis show that stochastic search also affords good results regarding diversity and rate of convergence, respectively. In the future, we would like to further improve ranking relevance and explore machine learning methods to personalize user's search experience.

\bibliographystyle{abbrv}
\bibliography{references}

\begin{thebibliography}{10}

\bibitem{Agrawal09}
R.~Agrawal, S.~Gollapudi, A.~Halverson, and S.~Ieong.
\newblock Diversifying search results.
\newblock In {\em Proceedings of the {Second} {ACM} {International}
  {Conference} on {Web} {Search} and {Data} {Mining} - {WSDM} '09}, page~5,
  Barcelona, Spain, 2009. ACM Press.

\bibitem{burges_ranknet_nodate}
C.~J.~C. Burges.
\newblock From {RankNet} to {LambdaRank} to {LambdaMART}: {An} {Overview}.
\newblock page~19, 2010.

\bibitem{chang_libsvm:_2011}
C.-C. Chang and C.-J. Lin.
\newblock {LIBSVM}: {A} library for support vector machines.
\newblock {\em ACM Transactions on Intelligent Systems and Technology},
  2(3):1--27, Apr. 2011.

\bibitem{Clarke08}
C.~L. Clarke, M.~Kolla, G.~V. Cormack, O.~Vechtomova, A.~Ashkan, S.~Büttcher,
  and I.~MacKinnon.
\newblock Novelty and diversity in information retrieval evaluation.
\newblock In {\em Proceedings of the 31st annual international {ACM} {SIGIR}
  conference on {Research} and development in information retrieval - {SIGIR}
  '08}, page 659, Singapore, Singapore, 2008. ACM Press.

\bibitem{cossock_subset_2006}
D.~Cossock and T.~Zhang.
\newblock Subset {Ranking} {Using} {Regression}.
\newblock In {\em Proceedings of the 19th {Annual} {Conference} on {Learning}
  {Theory}}, {COLT}'06, pages 605--619, Berlin, Heidelberg, 2006.
  Springer-Verlag.

\bibitem{dang_lemur_nodate}
V.~Dang.
\newblock The {Lemur} {Project}-{Wiki}-{RankLib}., 2013.

\bibitem{guan_ictnet_nodate}
F.~Guan, S.~Zhang, C.~Liu, X.~Yu, Y.~Liu, and X.~Cheng.
\newblock {ICTNET} at {Federated} {Web} {Search} {Track} 2014.
\newblock page~5, 2014.

\bibitem{karimzadehgan_stochastic_2011-1}
M.~Karimzadehgan, W.~Li, R.~Zhang, and J.~Mao.
\newblock A stochastic learning-to-rank algorithm and its application to
  contextual advertising.
\newblock In {\em Proceedings of the 20th international conference on {World}
  wide web - {WWW} '11}, page 377, Hyderabad, India, 2011. ACM Press.

\bibitem{Krestel2012}
R.~Krestel and P.~Fankhauser.
\newblock Reranking web search results for diversity.
\newblock {\em Information Retrieval}, 15(5):458--477, Oct 2012.

\bibitem{larkey_collection_2000}
L.~S. Larkey, M.~E. Connell, and J.~Callan.
\newblock Collection {Selection} and {Results} {Merging} with {Topically}
  {Organized} {U}.{S}. {Patents} and {TREC} {Data}.
\newblock In {\em Proceedings of the {Ninth} {International} {Conference} on
  {Information} and {Knowledge} {Management}}, {CIKM} '00, pages 282--289, New
  York, NY, USA, 2000. ACM.

\bibitem{lillis_probfuse:_2006}
D.~Lillis, F.~Toolan, R.~Collier, and J.~Dunnion.
\newblock {ProbFuse}: {A} {Probabilistic} {Approach} to {Data} {Fusion}.
\newblock {\em Proceedings of the 29th annual international ACM SIGIR
  conference on Research and development in information retrieval - SIGIR '06},
  page 139, 2006.
\newblock arXiv: 1409.8518.

\bibitem{mckinnon_convergence_1998}
K.~I.~M. McKinnon.
\newblock Convergence of the {Nelder}--{Mead} {Simplex} {Method} to a
  {Nonstationary} {Point}.
\newblock {\em SIAM Journal on Optimization}, 9(1):148--158, Jan. 1998.

\bibitem{palakodety_query_nodate}
S.~Palakodety and J.~Callan.
\newblock Query {Transformations} for {Result} {Merging}.
\newblock page~5, 2014.

\bibitem{powell_impact_2000}
A.~L. Powell, J.~C. French, J.~Callan, M.~Connell, and C.~L. Viles.
\newblock The {Impact} of {Database} {Selection} on {Distributed} {Searching}.
\newblock In {\em Proceedings of the 23rd {Annual} {International} {ACM}
  {SIGIR} {Conference} on {Research} and {Development} in {Information}
  {Retrieval}}, {SIGIR} '00, pages 232--239, New York, NY, USA, 2000. ACM.

\bibitem{powell_search_1973}
M.~J.~D. Powell.
\newblock On search directions for minimization algorithms.
\newblock {\em Mathematical Programming}, 4(1):193--201, Dec. 1973.

\bibitem{Radlinski06}
F.~Radlinski and S.~Dumais.
\newblock Improving personalized web search using result diversification.
\newblock In {\em 29th annual international ACM SIGIR conference}, pages
  691--692. ACM, Aug 2006.

\bibitem{Shannon:1949}
C.~E. Shannon.
\newblock A {Mathematical} {Theory} of {Communication}.
\newblock page~55, 1949.

\bibitem{sheldon_lambdamerge:_2011}
D.~Sheldon, M.~Shokouhi, M.~Szummer, and N.~Craswell.
\newblock {LambdaMerge}: {Merging} the {Results} of {Query} {Reformulations}.
\newblock In {\em Proceedings of the {Fourth} {ACM} {International}
  {Conference} on {Web} {Search} and {Data} {Mining}}, {WSDM} '11, pages
  795--804, New York, NY, USA, 2011. ACM.

\bibitem{shokouhi_robust_2009}
M.~Shokouhi and J.~Zobel.
\newblock Robust result merging using sample-based score estimates.
\newblock {\em ACM Transactions on Information Systems}, 27(3):1--29, May 2009.

\bibitem{si_using_2002}
L.~Si and J.~Callan.
\newblock Using {Sampled} {Data} and {Regression} to {Merge} {Search} {Engine}
  {Results}.
\newblock {\em SIGIR}, 8, 2002.

\bibitem{vogt_fusion_1999}
C.~C. Vogt.
\newblock Fusion {Via} a {Linear} {Combination} of {Scores}.
\newblock {\em Information Retrieval}, 1:151--173, 1999.

\bibitem{wu_adapting_2010}
Q.~Wu, C.~J.~C. Burges, K.~M. Svore, and J.~Gao.
\newblock Adapting boosting for information retrieval measures.
\newblock {\em Information Retrieval}, 13(3):254--270, June 2010.

\bibitem{zhai03}
C.~Zhai, W.~W. Cohen, and J.~Lafferty.
\newblock Beyond independent relevance: methods and evaluation metrics for
  subtopic retrieval.
\newblock In {\em 26th Annual International ACM SIGIR Conference}, SIGIR '03,
  pages 10--17, 2003.

\end{thebibliography}

%
\newpage
\appendix

\section{Evaluation of Optimal Entropy Solution}
\label{append:proof}
\begin{figure*}
    \centering
	\includegraphics[scale = 0.52]{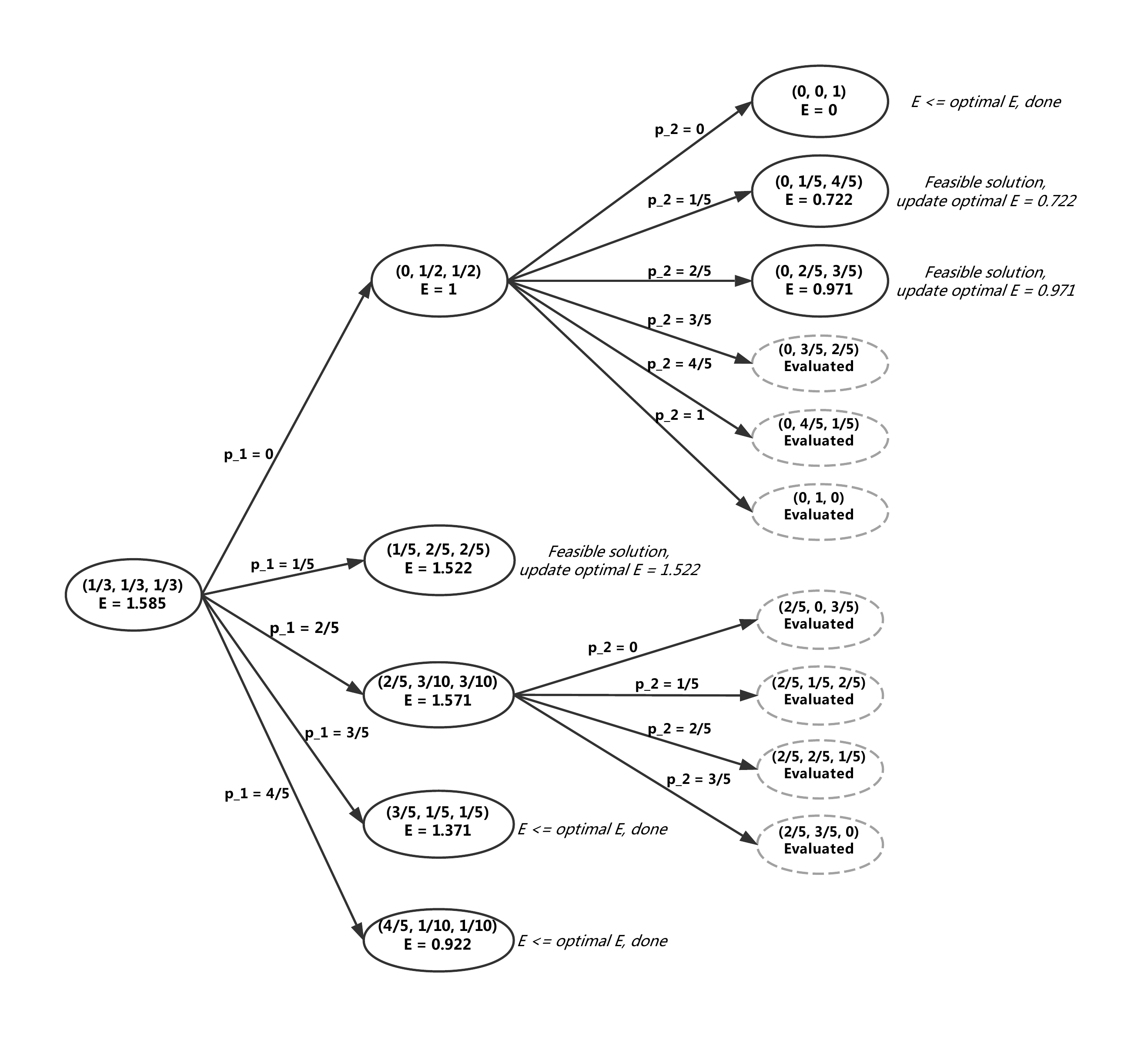}
	\caption{Calculate ideal Entropy by branch-and-bound algorithm}
	\label{illu} 
\end{figure*}
We use branch and bound to evaluate the optimal solution in Claim~\ref{ieproof2}. We still use the previous example of finding maximum entropy of 3 record types for top 5 documents. The question could be formatted as a linear programming problem:
\begin{gather}
\max \quad -[p_1\log{p_1} + p_2\log{p_2} + p_3\log{p_3}]  \\
\begin{aligned}
\textup{s.t.}\quad &p_1 + p_2 + p_3  =  1   \\
&p_i \in \{0,\frac{1}{5},\frac{2}{5},\frac{3}{5},\frac{4}{5},1\} \nonumber \\
\end{aligned}
\label{original}
\end{gather}
Its relaxation problem is as follows:
\begin{gather}
\max \quad -[p_1\log{p_1} + p_2\log{p_2} + p_3\log{p_3}]  \\
\begin{aligned}
\textup{s.t.}\quad &p_1 + p_2 + p_3  =  1  \nonumber\\
&p_i \geq 0\\
\end{aligned}
\label{relaxed}
\end{gather}
We have the following Claim:\\
\begin{claim}
The optimal entropy of the relaxation problem is always greater than or equal to that of the original problem. 
\label{relation}
\end{claim}
We will not go through a rigid proof for Claim~\ref{relation}.  The solutions of the original problem are always feasible ones to its relaxation problem, but not vice visa. Thus the optimal entropy of the relaxation problem is always an upper bound to that of the original problem. 

Making use of Claim~\ref{relation}, we apply branch and bound as follows:
\begin{enumerate}
\item We initialize the current maximum entropy to 0. We will update it whenever we find a feasible solution that generates a larger entropy than current one. 
\item For each sub problem, we check the optimal solution of its relaxation problem. 
\begin{enumerate}
\item If the optimal solution for the relaxation problem is feasible to the original problem, we do not need to evaluate solutions along this branch. As according to Claim~\ref{relation}, this optimal entropy is the best solution we could achieve along this branch. Furthermore, if the optimal entropy is larger than the current maximum entropy, we need to update the current maximum entropy. 
\item If the optimal solution for the relaxation problem is not feasible to the original problem, and its entropy is smaller than the current maximum entropy, then we do not need to evaluate solutions along this branch neither. The reason is that the optimal entropy of the relaxation problem is an upper bound of that of the original problem, as indicated in Claim~\ref{relation}, then all possible solutions along this branch will be bounded up by the optimal solution of the relaxation problem, therefore smaller than the current best entropy.
\end{enumerate}
\end{enumerate}

Following the above procedure, we initialize the current maximum entropy to 0. Then we calculate the optimal solution at $(1/3, 1/3, 1/3)$ with $E = 1.585$ for the relaxation problem (6). Because the optimal solution is not feasible of the original problem (5), we need to go through different branches.\\
\textbf{Branch}: 

\renewcommand{\labelenumii}{\theenumii}
\renewcommand{\theenumii}{\theenumi.\arabic{enumii}.}

1.  $p_1 = 0$. The problem becomes

\begin{gather*}
\max \quad - p_2\log{p_2} -p_3\log{p_3}  \\
\begin{aligned}
\textup{s.t.}\quad &p_2 + p_3  =  1\\
&p_i \in \{0,\frac{1}{5},\frac{2}{5},\frac{3}{5},\frac{4}{5},1\}\\
\end{aligned}
\end{gather*}

The optimal solution of the LP relaxation is at $(0, 1/2, 1/2)$ with $E = 1$. This is because when $p_0$ is fixed, the maximum entropy is achieved when $p2$ and $p3$ are the same. Since this optimal solution is not feasible to our original problem, we continue to branch along this direction.

1.1 $p_2 = 0$. The optimal solution is $(0,0,1)$ with $E = 0$. This is a feasible solution, but the entropy is not better than the current optimal entropy, which is 0.

1.2 $p_2 = 1/5$. The optimal solution is $(0,1/5,4/5)$ with $E =0.722$. It is a feasible solution and larger than current best entropy. Thus we need to update optimal entropy to 0.722.

1.3 $p_2 = 2/5$. The optimal solution is $(0, 2/5, 3/5)$ with $E=0.971$. It is a feasible solution and larger than current best entropy. Thus we need to update optimal entropy to 0.971.

1.4 $p_2 = 3/5$. The solution $(0, 3/5, 2/5)$ is evaluated previously as $(0, 2/5, 3/5)$ in case 1.3. Note that the order of the probability distribution does not affect entropy value. 

1.5 $p_2 = 4/5$. The solution $(0, 4/5, 1/5)$ is evaluated previously.

1.6 $p_2 = 1$. The solution $(0, 1, 0)$ is evaluated previously.

2. $p_1 = 1/5$. The optimal solution of the LP relaxation

\begin{gather*}
\max \quad -[\frac{1}{5}\log{\frac{1}{5}} + p_2\log{p_2} + p_3\log{p_3}]  \\
\begin{aligned}
\textup{s.t.}\quad &\frac{1}{5} + p_2 + p_3  =  1\\
&p_i \geq 0\\
\end{aligned}
\end{gather*}

is at $(1/5, 2/5, 2/5)$ with $E=1.522$. It is a feasible solution and larger than current best entropy. Thus we need to update optimal entropy to 1.522.

3. $p_1 = 2/5$. The optimal solution of the LP relaxation

\begin{gather*}
\max \quad -[\frac{2}{5}\log{\frac{2}{5}} + p_2\log{p_2} + p_3\log{p_3}]  \\
\begin{aligned}
\textup{s.t.}\quad &\frac{2}{5} + p_2 + p_3  =  1\\
&p_i \geq 0\\
\end{aligned}
\end{gather*}

is at $(2/5, 3/10, 3/10)$ with $E=1.571$. This is not a feasible solution, but its entropy is larger than current one, thus we continue to branch. 

3.1 $p_2 = 0$. The solution $(2/5, 0, 3/5)$ is evaluated previously.

3.2 $p_2 = 1/5$. The solution $(2/5, 1/5, 2/5)$ is evaluated previously. 

3.3 $p_2 = 2/5$. The solution $(2/5, 2/5, 1/5)$ is evaluated previously.

3.4 $p_2 = 3/5$. The solution $(2/5, 3/5, 0)$ is evaluated previously.

4. $p_1 = 3/5$. The solution of the LP relaxation

\begin{gather*}
\max \quad -[\frac{3}{5}\log{\frac{3}{5}} + p_2\log{p_2} + p_3\log{p_3}]  \\
\begin{aligned}
\textup{s.t.}\quad &\frac{3}{5} + p_2 + p_3  =  1\\
&p_i \geq 0\\
\end{aligned}
\end{gather*}

is at $(3/5, 1/5, 1/5)$ with $E=1.371$. The entropy is smaller than current optimal one, thus there is no need to branch along this direction.

5. $p_1 = 4/5$. The optimal solution of the LP relaxation

\begin{gather*}
\max \quad -[\frac{4}{5}\log{\frac{4}{5}} + p_2\log{p_2} + p_3\log{p_3}]  \\
\begin{aligned}
\textup{s.t.}\quad &\frac{4}{5} + p_2 + p_3  =  1\\
&p_i \geq 0\\
\end{aligned}
\end{gather*}

is at $(4/5, 1/10, 1/10)$ with $E=0.922$. The entropy is smaller than current optimal one, thus there is no need to branch along this direction.

Therefore, the solution $(1/5, 2/5, 2/5)$ with $E = 1.522$ is the optimal solution. Our method for calculating maximum entropy in CLAIM~\ref{ieproof2} finds the same result. To evaluate the maximum entropy for any given number of record types at certain position, we could always use branch and bound algorithm to verify that our method for maximum entropy calculation is correct. Figure~\ref{illu} illustrates how branch of bound is used to find the maximum entropy for this specific example. 

We conclude that we do not need to further branch a node in the graph when either one of the following conditions is satisfied:
\begin{enumerate}
\item The solution has been evaluated before. Here we make use of the property that the order of the probability distribution does not affect entropy value to cut many unnecessary branches. 
\item The optimal entropy of the relaxation problem is a feasible solution. This means that we find the optimal entropy along this branch. We just need to compare this value with the best solution we know.
\item The relaxation problem has an optimal entropy that is worse than the current best entropy. In this case, all possible solutions along this branch will be worse than the current best one.
\end{enumerate}

\section{Online Resources}
We will consider individual requests to access the source code. 

\end{document}